# Gold nanorods targeting tumor vasculature: Monte Carlo evaluation of the radioenhancement effects

Ali Taheri [a,*], Hans Rabus [b], Mayeen Uddin Khandaker [a,c,d], Farhad Moradi [e], David Andrew Bradley [a,f]

[a] Applied Physics and Radiation Technologies Group, CCDCU, Faculty of Engineering and Technology, Sunway University, 47500 Bandar Sunway, Selangor, Malaysia
[b] Physikalisch-Technische Bundesanstalt (PTB), 10587 Berlin, Germany
[c] The Miyan Research Institute, International University of Business Agriculture and Technology, 4, Embankment Drive Road, Sector-10, Uttara Model Town, Dhaka-1230, Bangladesh
[d] Department of Physics, College of Science, Korea University, 145 Anam-ro, Seongbuk-gu, Seoul, 02841, Republic of Korea
[e] Department of Biomedical Imaging, Faculty of Medicine, Universiti Malaya, Kuala Lumpur, Malaysia
[f] School of Mathematics and Physics, University of Surrey, Guildford, GU27XH, United Kingdom

* Corresponding author, email: alitmedical@gmail.com.

**Abstract**

Gold nanoparticles targeted to tumor vasculature can raise the dose to endothelial cells, and the enhancement depends on their position relative to the vessel wall. Despite growing interest in endothelial radiosensitization, intravascular and immediate perivascular dose enhancement has not been resolved at the level of individual endothelial cells for different nanoparticle locations. Earlier vascular dosimetry scored thin wall layers, analytic endothelial slabs, or tissue-region averages. This study quantifies the dose enhancement ratio (DER) of gold nanorods (AuNRs) in individual endothelial cells and in the perivascular region for three AuNR locations. Track-structure Monte Carlo simulations in TOPAS modeled a 10 µm-long, 16 µm-diameter microvessel segmented into six endothelial cells. AuNRs were placed homogeneously in the lumen, on the inner vessel wall (vascular-targeted), or on the outer vessel wall (extravasated) at 5–50 mg Au/g water within the vessel volume under 50 and 100 kVp X-rays. All DERs were corrected for the lack of charged-particle equilibrium (CPE) inherent to microscale simulations. Inner wall-attached AuNRs produced the highest microvessel-averaged DERs, reaching ~3.8 at 50 mg Au/g water and ~1.96 at 18 mg Au/g water under 50 kVp. The DER increased linearly with AuNR concentration and was higher at 50 kVp than at 100 kVp. Including AuNRs outside the simulated segment (infinite vessel approximation) increased endothelial DERs by 5–10%. In the perivascular region, the DER was highest within the first ~10 µm, where outer wall-attached AuNRs gave the highest values. The results set the physical dose enhancement foundation for endothelial and perivascular cells, guiding the design of vascular targeting radiosensitization.

## 1 Introduction

Nanoparticle radioenhancement represents a promising strategy to enhance the efficacy of radiation therapy by increasing the differential response between tumor and normal tissues (Gerken et al., 2023). The effectiveness of nanoparticle radioenhancers depends significantly on their localization within the tumor microenvironment. Targeting strategies are designed to optimize distribution and control cellular internalization through passive and active mechanisms, utilizing appropriate surface functionalization (Arcos Rosero et al., 2024; Moradi et al., 2021). While various targeting approaches exist based on extracellular, cellular, or subcellular compartments, vascular targeting has emerged as a particularly promising strategy (Peng et al., 2025).

Vascular disruptive agents (VDAs) are compounds that represent an extracellular targeting approach, directing therapeutic intervention toward tumor vasculature rather than the tumor cells themselves (Gómez-Escudero et al., 2025; Mita et al., 2013). This approach utilizes the abnormal features of tumor vasculature (Fu et al., 2026), including chaotic architecture, irregular branching, fragile and hyper-permeable vessels with enlarged intercellular gaps, and the overexpression of receptors such as integrin αvβ3 and vascular endothelial growth factor receptors (VEGFRs), which are minimally expressed in normal vessels (Graf et al., 2012; Lorenc et al., 2025; Sun et al., 2024). Compared with direct tumor cell or nuclear targeting, vascular targeting offers better accessibility and a potential amplification effect at the tissue level (Awad et al., 2023). However, this approach can also cause regions of hypoxia due to inadequate blood flow and oxygen delivery, which may, in turn, promote radioresistance and reduce the efficacy of conventional radiation therapy (Telarovic et al., 2021).

Targeting the tumor vasculature using nanoparticles can be achieved through both passive and active approaches. Passive targeting utilizes the enhanced permeability and retention (EPR) effect, where nanoparticles (typically 20-200 nm) accumulate in tumors due to leaky vasculature and poor lymphatic drainage. This allows nanoparticles to extravasate and concentrate in the perivascular region, the area immediately surrounding blood vessels, which is particularly important as this region can host therapy-resistant cancer stem cells and metastatic dispersion (Nowosad et al., 2023). In glioma, stem-like cells in the perivascular niche display radiation resistance (Pietras et al., 2014). Nanoparticle size critically influences the spatial distribution of nanoparticles, with larger particles remaining near the vascular system while smaller ones penetrate deeper into the tumor matrix (Perrault et al., 2009). Active targeting, on the other hand, is achieved through surface functionalization with ligands that bind to overexpressed endothelial receptors, such as Arginylglycylaspartic acid (RGD) targeting integrin αvβ3, anti-VEGFR antibodies, or the conjugation of VDAs and anti-angiogenic compounds (Kargozar et al., 2020). These mechanisms enable specific localization at the tumor vasculature while minimizing accumulation in normal vessels.

While the literature on nanoparticle radioenhancement (or nanoparticle radiosensitization) is dominantly focused on passive targeting mechanisms as well as cellular and nuclear targeting strategies (Schuemann et al., 2020; Taheri et al., 2023), vascular targeting presents convincing advantages that necessitate further investigation. This approach addresses limitations of inadequate tumor cell diffusion and poor tissue penetration by providing direct intravenous access to therapeutic targets without confronting physiological barriers (Jahanban-Esfahlan et al., 2018). Tumor endothelial cells exhibit more reliable and consistent receptor expression compared to highly variable tumor cellular targets, which depend on factors such as tumor type, origin, stage, and mutational characteristics. Furthermore, endothelial damage can be more efficient than clonogenic cell death, potentially affecting large tumor areas and suppressing metastasis (Garcia-Barros et al., 2003; Ngwa et al., 2010).

Recent experimental and computational studies have validated the therapeutic potential of vascular-targeted nanoparticle radioenhancement, establishing both proof-of-concept and mechanistic understanding. Several pioneering experimental investigations demonstrated significant therapeutic promise. Joh et al. investigated PEGylated gold nanoparticles (AuNPs) in glioblastoma multiforme cells and brain endothelial cells using 150-175 kVp X-rays, reporting increased endothelial DNA damage (sensitizer enhancement ratio of ~1.3) and improved mouse survival, confirming enhanced radiosensitivity of both tumor cells and associated vasculature (Joh et al., 2013). Yang et al. employed RGD-labeled AuNP probes of varying sizes (29, 51, and 80 nm) with Gd and $^{99m}$Tc labels. The targeted probes demonstrated 2- to 4-fold increases in intracellular gold uptake compared to non-targeted controls, with 80 nm NPs showing the greatest cellular uptake *in vitro*. However, for *in vivo* radiosensitization under $^{137}$Cs γ-ray irradiation (662 keV), the 29 nm particles proved most effective for tumor therapy (Yang et al., 2016).

Analytic and Monte Carlo (MC) calculations have provided quantitative predictions for vascular radioenhancement effects. Ngwa et al. calculated the endothelial dose enhancement analytically in tumor vasculature, comparing multiple brachytherapy sources ($^{103}$Pd, $^{125}$I, $^{169}$Yb) with 50 kVp X-rays and demonstrating that lower energy sources yielded the highest endothelial dose enhancement (Ngwa et al., 2010). The corresponding dose enhancement ratios (DERs, defined as dose with nanoparticle to dose without nanoparticles) were up to ~16 due to photoelectron and Auger electron contributions. Berbeco et al. calculated analytically the endothelial DER values of 1.2-4.4 for AuNPs attached to endothelium under 6-MV photon irradiation for intravascular AuNP concentrations of 7 to 140 mg Au/g water (Berbeco et al., 2011). Lin et al. used TOPAS simulations to assess vascular dose enhancement across multiple radiation sources, reporting an additional dose of up to 207% of the prescribed dose (DER of ~3.07) at vascular walls with wall-attached AuNPs under 250 kVp photons, and highlighting the importance of nanoparticle placement and energy selection (Lin et al., 2015). Mansouri et al. compared five high-Z nanoparticles (gold, bismuth, gadolinium, iridium, and hafnium) using mathematical modeling and Geant4 simulations, revealing material- and energy-dependent DERs that reached up to 14 (Mansouri et al., 2021). Díaz-Galindo and Garnica-Garza used the PENELOPE MC code in a 1 mm$^3$ branching vascular model with 500 nm voxels and 10 mg Au/g. For 100 keV and 220 kVp photons, the average dose in tumor tissue was 2 to 3 times higher with NPs in the bloodstream or on the inner vessel walls than with nanoparticles in the tissue. For 6 MV photons the ratio was approximately 1 (Díaz-Galindo and Garnica-Garza, 2024).

Despite these significant contributions, existing studies carry methodological limitations that obstruct a comprehensive understanding of vascular radioenhancement mechanisms. In particular, dose in individual endothelial cells was not reported for AuNRs at different locations in the same vessel. In addition, these studies avoided the lack of secondary charged-particle equilibrium (CPE) by using analytic models, prescribed reference dose, or larger phantoms to establish equilibrium. A simulation that transports a confined beam through explicit nanoparticles and scores micrometer-scale volumes does not satisfy CPE. Without such correction, the dose without nanoparticles is underestimated and the DER is overestimated (Rabus et al., 2021b, 2019). Furthermore, accurate comparisons between different nanoparticle localizations within and around the vasculature remain absent. More critically, the literature has focused predominantly on vascular dose enhancement, while dose enhancement within the perivascular region, a biologically relevant compartment associated with radioresistance, has been largely overlooked.

Building upon our previous investigations on high-Z metallic nanorods for nanoparticle-mediated radiation therapy (Taheri et al., 2025a, 2025b, 2024), the present study addresses these gaps by examining how the localization and concentration of gold nanorods influence radioenhancement in individual endothelial cells and in the perivascular region. These investigations are summarized in a preprint letter (Taheri et al., 2026), the present study addresses these gaps by examining how the localization and concentration of gold nanorods within tumor vasculature govern radioenhancement outcomes, both at the level of individual endothelial cells and within the perivascular region surrounding the microvessel. Simulations are performed using the TOPAS (TOol for PArticle Simulation) track-structure Monte Carlo code (Perl et al., 2012), under 50 and 100 kVp X-ray spectra. Dose enhancement ratios within individual endothelial cells are calculated and subsequently corrected for charged-particle equilibrium to remove beam-size–related bias. In addition, radial dose enhancement profiles are evaluated from the vessel surface out to 40 μm into the surrounding tissue, enabling direct assessment of dose propagation beyond the vascular wall. These analyses are conducted for six AuNR concentrations spanning 5–50 mg Au/g water. By focusing on 50 and 100 kVp X-ray spectra, the clinical scope of this work is specifically oriented toward superficial and orthovoltage radiotherapy, providing dosimetric insights primarily applicable to shallow-seated malignancies.

## 2 Methodology

A multiscale Monte Carlo simulation approach was implemented to evaluate how the distribution and concentration of gold nanorods in tumor vasculature affect radioenhancement efficiency under X-ray irradiation. In the first stage, Phase Space Files (PSFs) for 50 and 100 kVp X-ray spectra were generated using the TOPAS code following previously reported frameworks (Taheri et al., 2024). The PSFs subsequently served as photon irradiation sources for the micro- and nanoscale simulations of tumor vessel irradiation under various gold nanorod distribution and concentration scenarios. In the microscopic simulations, a spatially confined photon beam matched to the vessel segment was employed to restrict particle transport to the region of interest and enable efficient cell-resolved dose scoring. This narrow beam setup is necessary in nanoparticle radiosensitization simulations, as modeling a laterally extended realistic irradiation field would require excessively large particle histories and computational power. However, such geometric confinement prevents the satisfaction of charged-particle equilibrium (CPE) at the micrometer scale, since secondary electrons generated outside the beam boundary do not fully contribute to the local energy deposition. The bias introduced by this lack of CPE was corrected here to estimate DER values under equilibrium conditions, and an additional correction was applied to account for dose contributions arising from photon interactions in AuNRs located outside the simulation geometry.

### 2.1 Phase space generation

The simulation geometry for the generation of the PSFs consisted of a 1×1 $cm^2$ beam directed at a 4×4×4 $cm^3$ water phantom from a source-to-surface distance of 5 cm (Figure 1a). The X-ray photon energy spectra were imported from SpekPy (Poludniowski et al., 2021) at peak voltages of 50 and 100 kVp, representing energy ranges commonly employed in superficial and orthovoltage radiotherapy for shallow-seated tumors such as skin carcinomas and melanomas (Han et al., 2022). The phase space scoring region was defined as a 100×100×1 $\mu m^3$ water slab positioned at a depth of 2 mm within the phantom, representing the typical depth of superficial tumor locations. For macroscale simulations the *g4em-standard_opt4* physics module was utilized for PSF generation, as it uses Livermore data library and provides the most accurate Geant4 electromagnetic models optimized for energies ranging from a few hundred eV to hundreds of GeV (Sakata et al., 2018). A typical secondary particle production threshold of approximately 100 eV was set to make a balance between computational efficiency and accuracy. With $10^9$ primary histories statistical uncertainties in the absorbed dose of less than 1% were achieved. The uncertainty is expressed as relative standard deviation, calculated as the standard deviation (STD) reported by TOPAS divided by the total scored dose value, expressed as a percentage.

### 2.2 Irradiation of the tumor vessel

For the subsequent micro- and nanoscale simulations, the *g4em-dna* physics module was used, which explicitly models track-structure interactions down to energies as low as a few eV (cut-off ~7.4 eV for electrons). This module incorporates detailed liquid water interaction models, which are crucial for radioenhancement and DNA damage simulations based on specialized Geant4-DNA physics processes rather than standard electromagnetic data libraries.

Following the geometry used in the literature (Lin et al., 2015; Mansouri et al., 2021), a tumor vessel section (10 µm length, 16 µm inner diameter, 2 µm wall thickness) was modeled within a 100×100×100 $\mu m^3$ water cube and exposed to X-ray irradiation from the generated PSFs (Figure 1b). The phase space was rescaled to a 20×10 $\mu m^2$ planar source to match the target vessel dimensions using the *PhaseSpaceScaleXPosBy* command in the TOPAS input file. For simulations involving nanorods distributed outside the vessel wall, the planar source was slightly extended to 20.3×10 $\mu m^2$, as described below. The vessel was divided into six segments, each representing one endothelial cell (approximately 10×10×2 $\mu m^3$), following the dimensions frequently used in previous simulation studies (Mansouri et al., 2021; Ngwa et al., 2010). These cell volumes were used to measure absorbed dose for different nanorod distribution and concentration scenarios. This idealized geometric assumption was deliberately chosen to provide a controlled, reproducible baseline for evaluating our CPE-correction methodology and radial dose, which would otherwise be convoluted by the extreme anatomical variability of patient-specific microvessels.

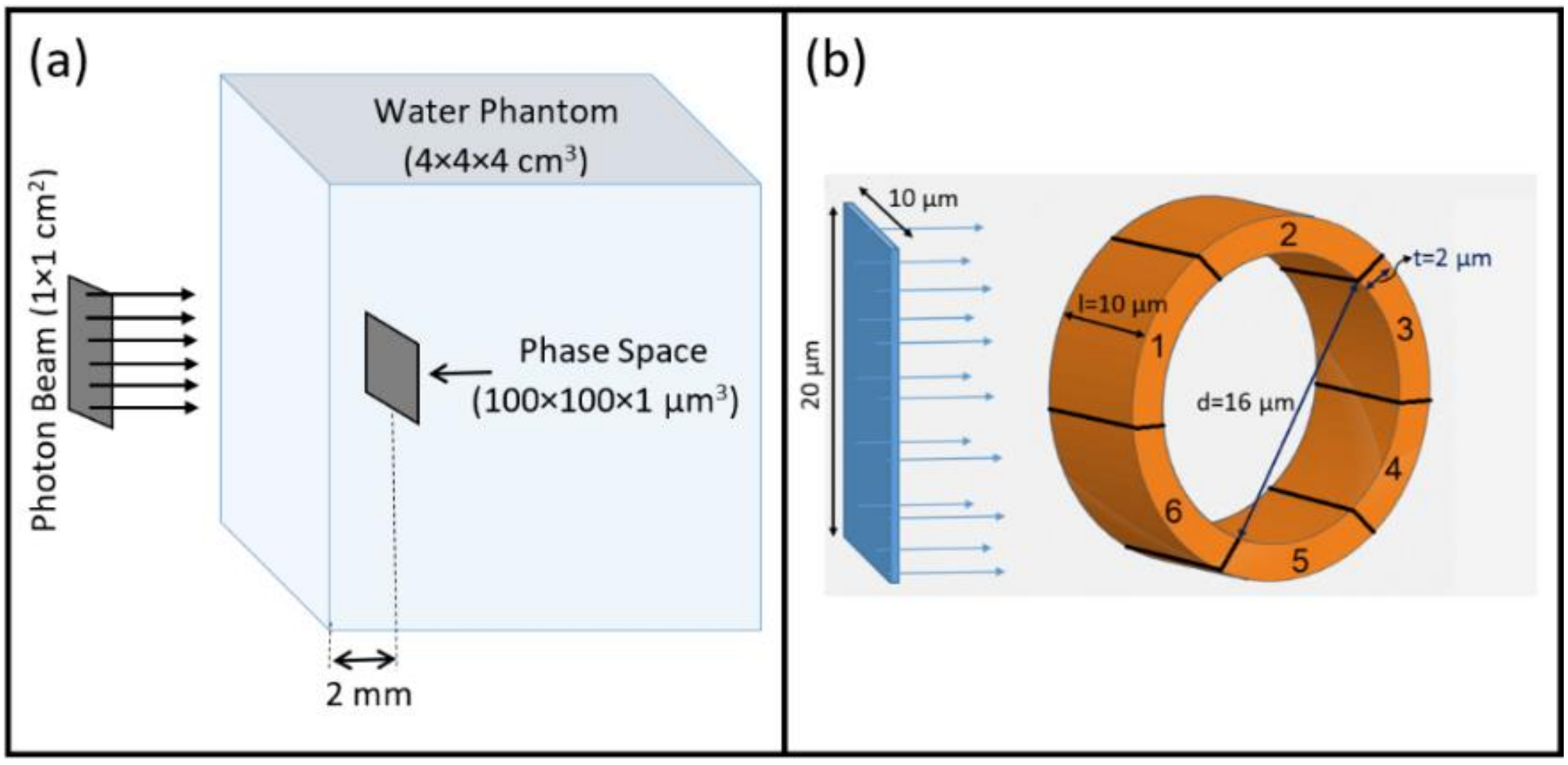

*Figure 1 Schematic illustration of the simulation setup: (a) Generating the initial phase space, (b) Photons originating from the rescaled phase space, with the field size matching the target tumor vessel dimension. The dimensions are not to scale and are for illustration purposes.*

Three distinct gold nanorod distribution scenarios were investigated as demonstrated in Figure 2:

i. **Homogeneous distribution** ("Homogen"): nanorods uniformly distributed within the vessel lumen, representing the initial distribution during intravenous circulation
ii. **Inner wall attachment** ("InnerWall"): nanorods attached to the vessel's internal surface, representing active targeting to tumor endothelial cells
iii. **Outer wall attachment** ("OuterWall"): nanorods positioned at the vessel's external surface, representing extravasation into the perivascular space

To generate these configurations, the *GenerateRandomCylinders* template from TOPAS-nBio extension (Schuemann et al., 2019) was employed to create the homogeneous nanorod distribution. Python scripts were then developed to modify this distribution and generate the InnerWall and OuterWall configurations. The nanorod centers were placed at random positions and orientations within a 50 nm thick region between 100 and 150 nm from the wall in both the InnerWall and OuterWall configurations. For the OuterWall configuration, the vertical beam size was therefore 20.3 µm instead of 20 µm to ensure that all nanorod centers were located in the beam.

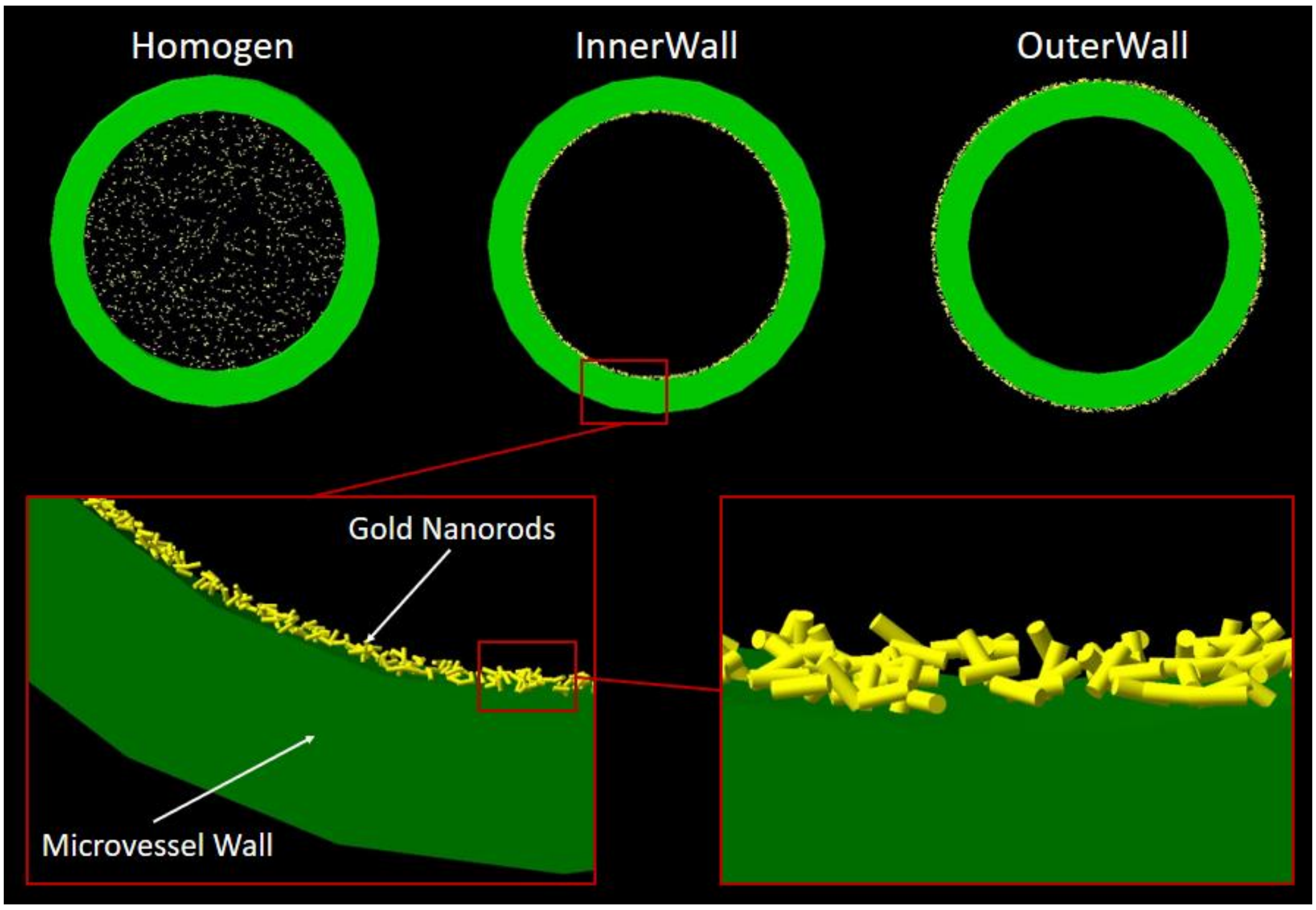

*Figure 2 Cross-sectional view of the three nanorod distributions, as well as the zoomed-in view of the InnerWall distribution, illustrating the random distribution of gold nanorods*

To evaluate the impact of nanorod concentration, different concentrations (expressed as milligrams of gold per gram of water - mg Au/g water) were compared for each distribution pattern. These concentrations were calculated based on the gold mass density (19.32 g/cm³) and the nanorod dimensions of 60 nm in diameter and 180 nm in length (aspect ratio = 3). Literature commonly reports a tumor-averaged gold concentration of approximately 7 mg Au/g water following systemic administration of gold nanoparticles. Higher concentrations have been reported at early time points, with values reaching ~18 mg Au/g water approximately 2 minutes after intravenous injection of 2.7 g Au/kg body weight in mice, corresponding to peak vascular confinement prior to significant extravasation (Cho et al., 2009; Hainfeld et al., 2008, 2004; Ngwa et al., 2010). These concentrations represent averages over the entire tumor volume, including both vascular and extravascular compartments. Since tumor vasculature can constitute approximately 5-50% of total tumor mass (Forster et al., 2017), conservation of gold mass implies that the local gold concentration within the vascular compartment may exceed the tumor-averaged value by a factor of approximately 2 to 20, depending on vascular density. Under this assumption, tumor-averaged concentrations in the range of 7 mg Au/g water correspond to effective vascular concentrations of approximately 14-140 mg Au/g water. To span a physiologically plausible and conservative part of this range, the present study investigated six gold concentrations (5, 7, 10, 18, 30, and 50 mg Au/g water) within the cylindrical volume containing the modeled microvessel (16 µm inner diameter). To eliminate bias due to the varying total gold mass, the same number of nanorods was used for all distributions.

For each combination of nanorod distribution and concentration, energies imparted per mass were measured within the six endothelial cells and compared to control conditions (without gold nanorods) to calculate DERs. Radial distributions of energy imparted per mass were also evaluated using concentric water cylindrical shells of 1 µm radial thickness and 10 µm length surrounding the vessel section, extending from the vessel's outer surface to 40 µm radially, to determine spatial DER profiles around the vessel.

### 2.3 Charged-Particle Equilibrium correction

In microscale Monte Carlo simulations of dosimetric effects of nanoparticles using confined beam geometries, the lack of long-range CPE is known to lead to a systematic overestimation of DERs (Moradi et al., 2022; Rabus et al., 2019, 2021a; Zygmanski et al., 2013). This overestimation arises because with a beam confined to the size of a small scoring volume, the energy imparted in the volume by electrons produced outside and entering it is underestimated since electrons produced in a macroscopic irradiation outside the region covered by the confined beam are neglected. Therefore, the dose obtained without nanoparticles is underestimated and so is the contribution to dose from electrons produced in photon interactions outside the scoring volume dose in the presence of high-Z nanoparticles. Proper correction for these non-equilibrium effects is therefore essential for obtaining absorbed dose and a physically meaningful DER estimation.

To account for the lack of CPE in microscale simulations, DERs were estimated by using a two-stage approach considering the ratios of dose and dose contributions to photon fluence. In the first stage, the macroscopic dose-per-fluence ratios under CPE conditions, $(D/\Phi)_{\mathrm{macro}}$, were estimated for the relevant beam qualities and depth using the calculation methods from (Rabus et al., 2019, 2020). The resulting values in 2 mm depth of water were $6.31\times10^{-13}$ Gy cm² for the 50 kVp and $4.17\times10^{-13}$ Gy cm² for the 100 kVp x-ray spectra used here. These ratios represent the equilibrium absorbed dose per fluence that would be obtained in an extended, laterally unconfined beam geometry (Rabus et al., 2021b).

In the second stage, the dose contributions arising from the confined beam were obtained from the results of the microscale Monte Carlo simulations of the endothelial cells and surrounding cylindrical shells, performed both in the absence and presence of AuNRs. The nanoparticle-induced excess dose per primary fluence, $(\Delta d/\Phi)_{\mathrm{micro}}$, was defined as the difference between the fluence-normalized doses scored with and without AuNRs. The CPE-corrected DER was then estimated as:

$$DER = 1 + (\Delta d/\Phi)_{micro} \;/\; (D/\Phi)_{\;macro} \qquad \textit{Eq. 1}$$

expressing the nanoparticle-induced dose enhancement relative to a macroscopic equilibrium dose and avoiding the DER overestimation caused by non-equilibrium conditions in confined beam geometries. The normalization steps, fluence reconstruction procedures, and intermediate quantities are provided in Supplementary Material 1. Because $(D/\Phi)_{\mathrm{macro}}$ is identical for all endothelial cells, the DER averaged over the six cells equals the DER obtained from the total dose scored in the six cells with and without AuNRs.

### 2.4 Estimate of long-range contributions from AuNRs

Apart from the lack of CPE in the absence of AuNRs, the microscopic simulations also do not account for dose contributions from nanorods in upstream and downstream sections of the vessel, i.e. adjacent to neighboring endothelial cells outside this section. In a macroscopic irradiation, electrons produced by photon interactions in these AuNRs can also contribute to the dose in the vessel segment considered in the microscopic simulation.

To estimate the magnitude of effect from this neglected dose contribution in the simulations with AuNRs, a development of the theoretical framework of (Rabus and Thomas, 2025) was used. In this approach, it was assumed that the blood vessel in a region around the segment shown in Figure 1 can be described by a cylinder of the same diameter. Assuming uniform concentrations in the regions containing AuNRs (lumen of the vessel or cylindrical shells at the inside or outside of the cylinder ring containing the cells), the radial distributions of AuNRs around the center of a cell nucleus and at points at distances between 1 µm and 20 µm from the outer wall of the cell ring were determined. This determination was based on analytical calculations of the proportion $\omega(r)$ of the solid angle for which the points on a spherical surface (radius $r$) around the considered points were inside the region containing AuNRs (Rabus and Thomas, 2025). This radial distribution of AuNRs was then used as weighting function for the radial distribution of additional dose $\Delta D(r)$ around the center of an AuNR in which an ionizing photon interaction occurred. From the integral of the weighted radial distribution of $\Delta D(r)$ around an AuNR undergoing an ionization, the DER resulting from the expectation of the dose contribution from AuNRs was then obtained by Eq. 2:

$$DER = 1 + n_{NR} \times N_{i,\Phi} \times \int 4\pi\omega(r)\Delta D(r) r^2 dr \;/\; (D/\Phi)_{macro} \qquad Eq.\ 2$$

Here, $N_{i,\Phi}$ is the expected number of photon interactions in an AuNR per fluence $\Phi$, and $n_{\mathrm{NR}}$ is the number density of AuNRs which is related to the concentration per mass $c_m$ by $n_{\mathrm{NR}} = c_m \times \rho_w / m_{NR}$ with $\rho_w$ the density of water and $m_{NR}$ the mass of an AuNR.

The radial distribution around an AuNR was approximated by that around a spherical gold nanoparticle of 100 nm diameter, which has approximately the same mass as the 60 nm × 180 nm AuNRs considered in this work. Equivalent volume nanospheres and nanorods differ in the propagation and absorption of secondary electrons (Milutinović et al., 2023), so this approximation introduces an error that was not quantified here. The corresponding data were also taken from Li et al. (Li et al., 2024). It must be noted, however, that the approach of (Rabus and Thomas, 2025) does not take into account the anisotropy of secondary electron emission and resulting potential differences in the dose received by different cells owing to their location along the photon beam and entailed changes in the photon spectrum. Therefore, these estimates can only be related to the dose enhancement averaged over cell positions. Moreover, blood flow dynamics were not incorporated into the present calculations. Intratumoral capillaries, particularly within the tumor core, are known to exhibit low red blood cell velocities on the order of 10-50 µm/s due to hypoperfusion and structural abnormalities (Kamoun et al., 2010). Energy deposition by ionizing radiation occurs on the time scale below pico-seconds so that the molecules and nanoparticles interacting with radiation move only by sub-atomar distances. Therefore, the static model employed in these simulations serves as a practical and accurate approximation for instantaneous energy deposition, successfully isolating fundamental radiation physics from complex flowing blood conditions.

The DER according to Eq. 2 was evaluated for points at distances from the outer surface of the vessel between 1 µm and 20 µm. In addition, it was also evaluated averaged over the nucleus of a cell in the vessel wall. This averaging of the DER is possible because the ratio of dose to fluence in the denominator can be assumed to be constant.

## 3 Results and Discussion

### 3.1 Cellular dose enhancement

The DER was evaluated in six endothelial cells for three nanorod distributions under 50 and 100 kVp X-rays at a reference concentration of 18 mg Au/g water. As shown in Figure 3, the DER shows a strong dependence on the spatial positioning of nanorods within the vasculature system. The attachment of nanorods to endothelial cells (e.g. via VDAs) significantly enhances the absorbed dose in these cells compared to nanorods dispersed throughout the bloodstream. The observed DER values range from 1.4 to 2.2 for the InnerWall nanoparticle distribution, which is representative of vascular targeting of gold nanorods under the two X-ray spectra. This enhancement effect is attributed primarily to the substantial contribution of low- and moderate-energy secondary Auger-Meitner electrons to physical dose enhancement, a phenomenon previously well-documented in the literature (Mansouri et al., 2021; Ngwa et al., 2010). Nonetheless, it should be noted that these comparisons are qualitative, as the present work reports physical DERs rather than biological endpoints (such as DNA damage), which involve additional biological radiosensitization mechanisms beyond physical dose enhancement.

The simulation study by Lin et al. (2015) reported a peak additional dose of up to 207% of the prescribed 30 Gy dose (DER of ~3.07) in a 10 nm layer at the inner wall, for wall-attached 10 nm AuNPs at 20 mg Au/g in an 8 µm vessel under 250 kVp irradiation. For a homogeneous distribution under the same conditions, the wall-averaged additional dose was 43% (DER of ~1.43) (Lin et al., 2015). The present results indicate a more moderate increase in DER compared to these peak values. As illustrated in Figure 4, the microvessel DER, averaged over the six endothelial cells, for the InnerWall and Homogen configurations are approximately 1.96 and 1.38 under 50 kVp, and 1.65 and 1.28 under 100 kVp, respectively. The discrepancy between our findings and those of Lin et al. (2015) is likely attributed to two key factors: differences in the X-ray spectra employed and methodological variations in NP distribution. In Lin et al., all nanoparticles were directly attached to the lumen wall without accounting for the more realistic scenario where NPs are coated with a functionalization layer, which effectively increases the distance between the NP and the cell, and decreases the delivered dose to the endothelial cell. Although the present simulations do not explicitly model the coating layer, this spatial association was taken into account by randomly placing nanorod centers between 100 and 150 nm from the wall, as stated in Section 2.2 (see Figure 2). The minimum center distance of 100 nm exceeds the nanorod half-length of 90 nm and therefore avoids any overlap between the nanorods and the microvessel wall. Since this approach more closely approximates real-world conditions, it yields lower DER values that may be regarded as a more precise representation of the actual dose delivered to the cells compared to previous studies.

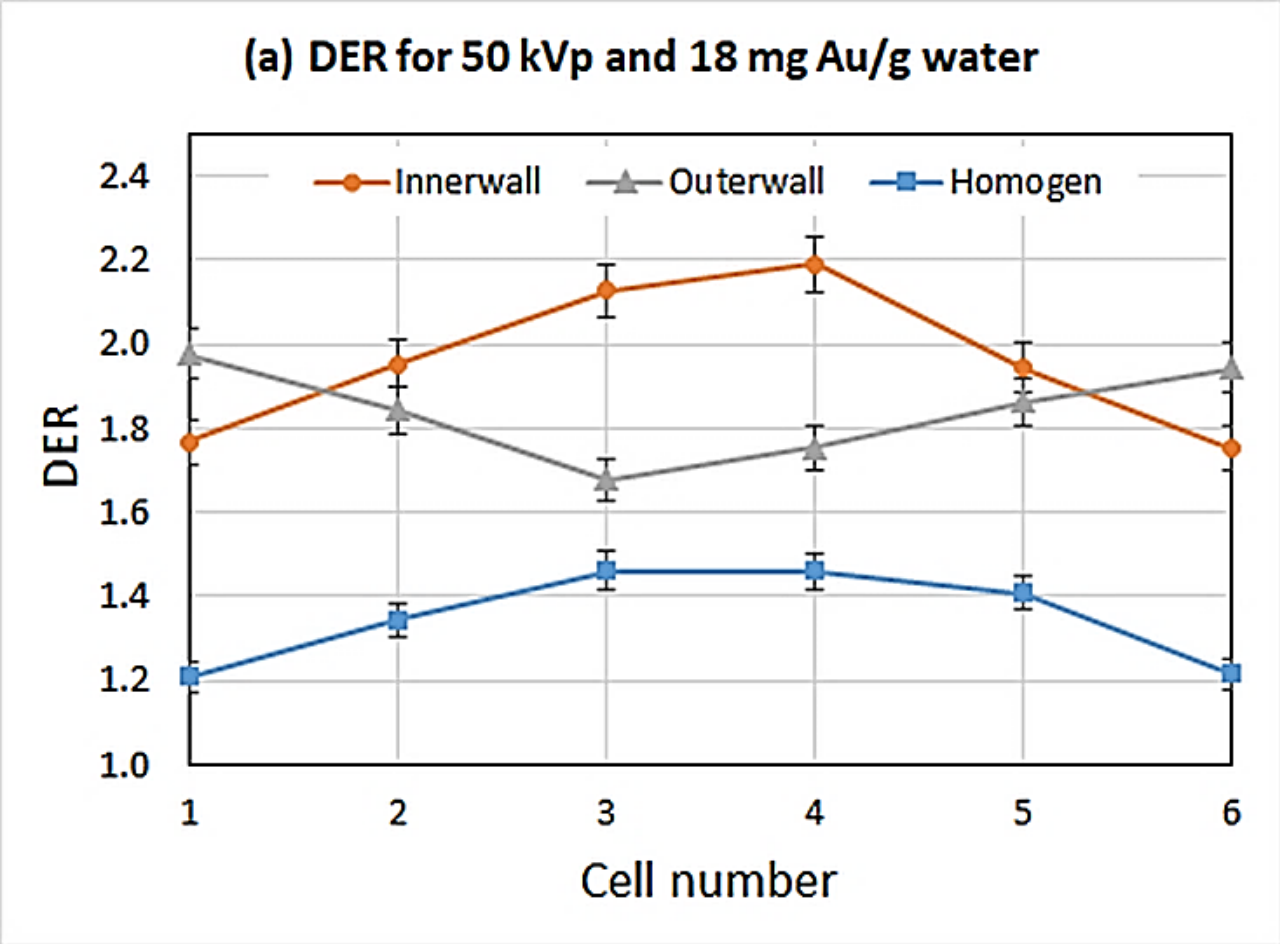

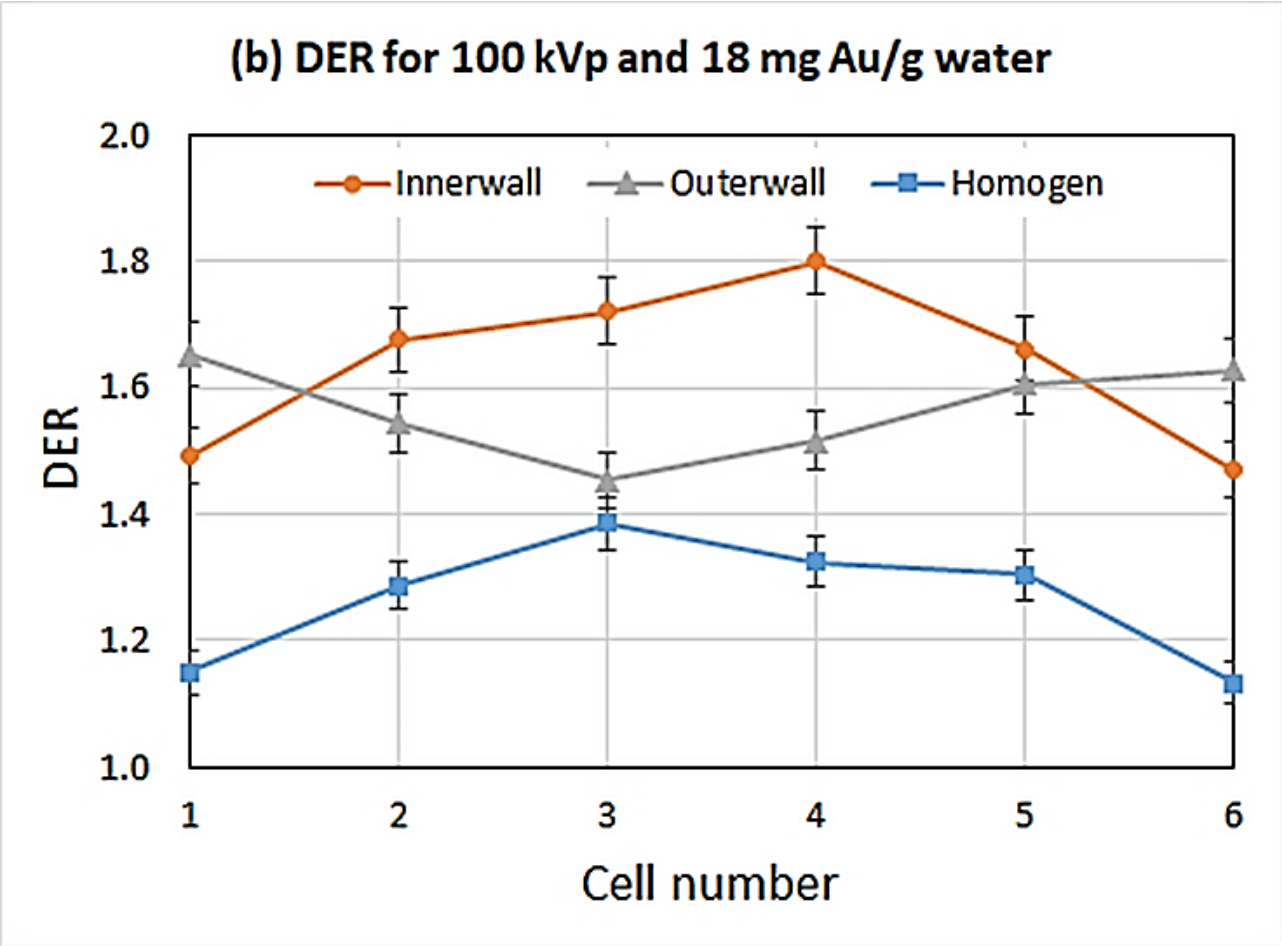

*Figure 3 DER in six endothelial cells for three examined nanorod distributions under (a) 50 kVp, and (b) 100 kVp X-ray irradiation. Error bars show the standard deviation (STD) reported by TOPAS.*

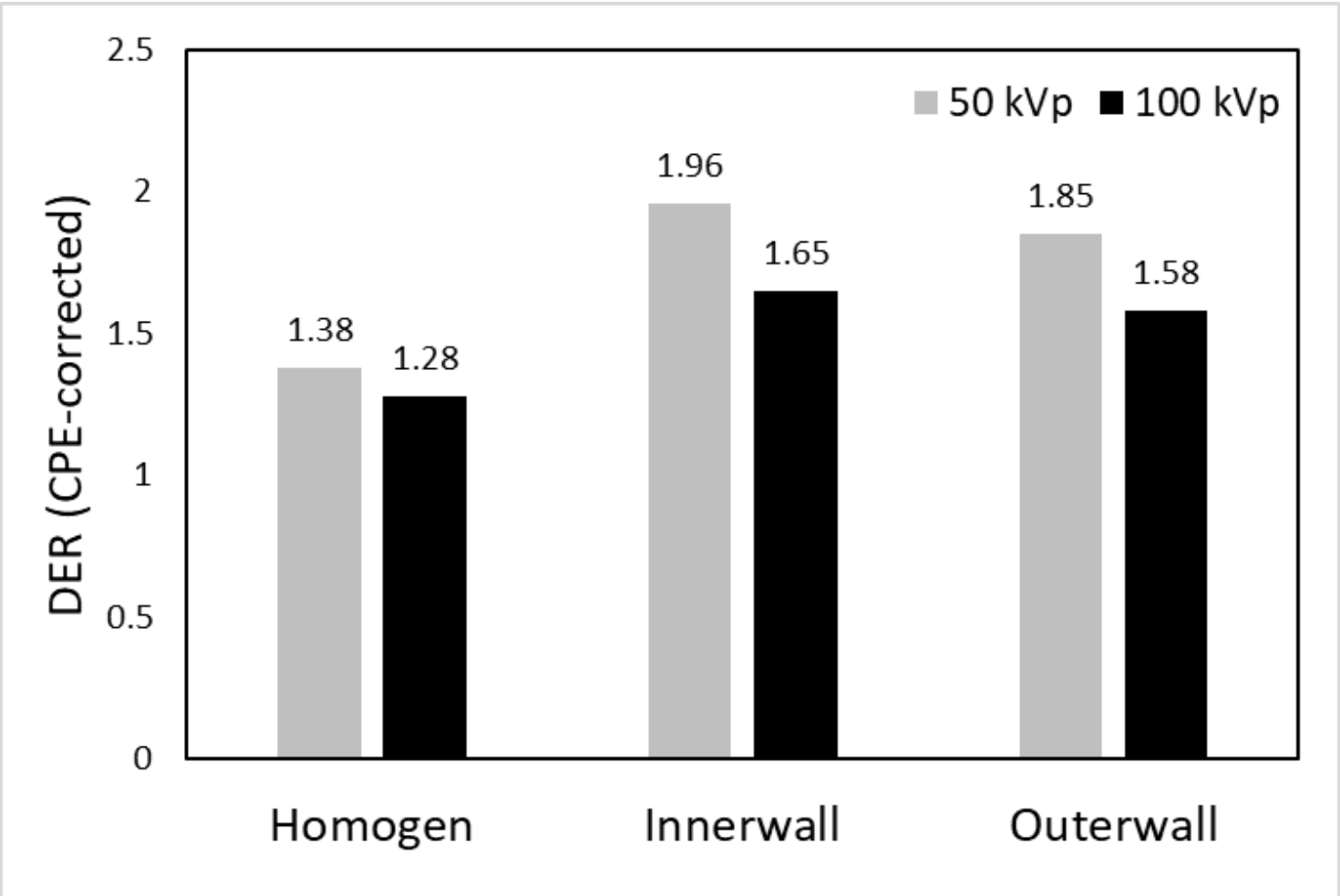

*Figure 4 DER averaged over the six endothelial cells (microvessel DER) for the three distributions of AuNRs at a concentration of 18 mg Au per g water under two X-ray spectra*

Table 1 summarizes the effect of including dose contributions from nanorods attached to adjacent endothelial cells on the average cell DERs. Accounting for neighboring nanorods increases the endothelial cell DERs by approximately 5-10% across all irradiation energies and configurations. This corresponds to an increase of the nanoparticle-induced enhancement ($\mathrm{DER} - 1$) by approximately 10-20% for the InnerWall and OuterWall distributions and by approximately 25-45% for the Homogen distribution. Importantly, applying this correction does not alter the overall trends observed in the uncorrected results. The infinite vessel segment continues to show higher DERs at 50 kVp compared to 100 kVp, and nanorods localized close to the vessel wall (InnerWall and OuterWall) consistently outperform the homogeneous lumen distribution. This confirms that the superiority of vascular-targeted nanorods is robust and not an artifact of neglecting intercellular dose contributions.

Furthermore, endothelial cell and nuclear geometries were simplified in the present model, whereas real tumor endothelial cells can be elongated, with cell lengths reaching several tens of micrometers and nuclei deviating from spherical shapes (Versaevel et al., 2012). However, nanoparticle-mediated dose enhancement is highly localized, with secondary electron ranges limited to a few micrometers. The doses reported here from the simulation geometry shown in Figure 1b therefore represent the dose delivered to a local nuclear sub-volume close to vascular-targeted nanorods rather than a whole-nucleus average. However, with an extended nuclear geometry (e.g., to a 50 µm-long cylindrical nucleus), each compartment of the nucleus would receive the same nanoparticle-induced dose as quantified in this study.

*Table 1 DER averaged over the six endothelial cells at 18 mg Au/g water for the simulated 10 µm vessel section and for an effectively infinite vessel (Section 2.4), for different nanoparticle distributions and X-ray energies. The ratio is the infinite-vessel DER divided by the 10 µm-section DER.*

| **Case** | **50kVp-Homogen** | **100kVp-Homogen** | **50kVp-InnerWall** | **100kVp-InnerWall** | **50kVp-OuterWall** | **100kVp-OuterWall** |
|---|---|---|---|---|---|---|
| 10 µm vessel section | 1.38 | 1.28 | 1.96 | 1.65 | 1.85 | 1.58 |
| Infinite vessel | 1.48 | 1.41 | 2.06 | 1.78 | 1.94 | 1.69 |
| Ratio | 1.07 | 1.10 | 1.05 | 1.08 | 1.05 | 1.07 |

Figure 5 shows the microvessel DER, averaged over six endothelial cells, as a function of gold nanorod concentration (5 - 50 mg Au/g water) for the examined scenarios, demonstrating that DER increases linearly with concentration, consistent with previous reports by Ngwa et al. (2010) for brachytherapy sources (Ngwa et al., 2010) and Berbeco et al. (2011) for megavoltage X-ray irradiation (Berbeco et al., 2011). The observed DER values lie between the enhancement ranges previously attributed to Auger electrons and photoelectrons by Ngwa et al. While their study showed a substantially larger contribution from Auger electrons compared with photoelectrons, the present results reflect the combined contribution of both electron types to the overall radioenhancement mechanism, in agreement with established expectations (Liu et al., 2015).

The dependence of dose enhancement on gold nanoparticle concentration can also be interpreted within the theoretical framework proposed by (Rabus et al., 2019, 2020). In this work, it was shown that for a homogeneous mixture of gold and water, the average DER depends on the gold mass fraction ($x_{Au}$), and the ratio of absorbed dose in gold to that in water ($D_{Au}/D_w$) under conditions of CPE. According to Eq. (8) in that work (Rabus et al., 2020, 2019), this relationship is given by

$$\overline{DER}_{mix} = 1 + x_{Au} \times (\frac{D_{Au}}{D_w} - 1) \qquad \textit{Eq. 3}$$

This expression explicitly demonstrates that the excess enhancement (DER - 1) scales proportionally with the gold mass fraction. The gold mass fraction $x_{Au}$ is related to the concentration $C_{Au}$ used in this work by $x_{Au} = C_{Au}/(1 + C_{Au})$ so that for small values of the gold concentration, mass fraction and concentration are essentially the same.

The DER for a mixture represents a theoretical upper limit for the average enhancement within a spherical volume surrounding a nanoparticle since a part of the energy transferred to electrons by a photon interaction in the nanoparticle is absorbed within it (Koger and Kirkby, 2016). This self-absorption increases with nanoparticle size and is largest for low-energy beams (Fernandes Alba et al., 2025). However, for nanoparticles the same proportionality between (DER - 1) and gold concentration is expected (Eq. 8 in (Rabus et al., 2021b)) as long as the nanoparticle concentration remains reasonably low. For larger nanoparticle concentrations, the 'sink effect' described by (Brivio et al., 2017), where secondary electrons produced outside the nanoparticles deposit significant proportion of their energy in them, may lead to non-linear dependence on gold concentration. Similar non-linearity may arise if the concentration of nanoparticles gets so high that a significant proportion of fluorescence photons produced in the de-excitation cascade after a photoabsorption is absorbed by other nanoparticles. In the cases of the nanoparticles attached to the inner or outer walls of the vessel, the concentration of gold in the region containing AuNRs is much larger than the average concentration inside the cylinder representing the vessel segment. In the case of 50 mg Au/g water average concentration, the gold mass fraction in the 50 nm thick region containing the AuNR centers is almost 80%. However, the concerns raised in (Rabus, 2024) on such high mass fractions do not apply to the present results since the AuNRs are not located inside the considered cell nucleus. Therefore, the present observations support the expected linear scaling of DER with gold concentration with a slope depending on the given radiation quality and material composition.

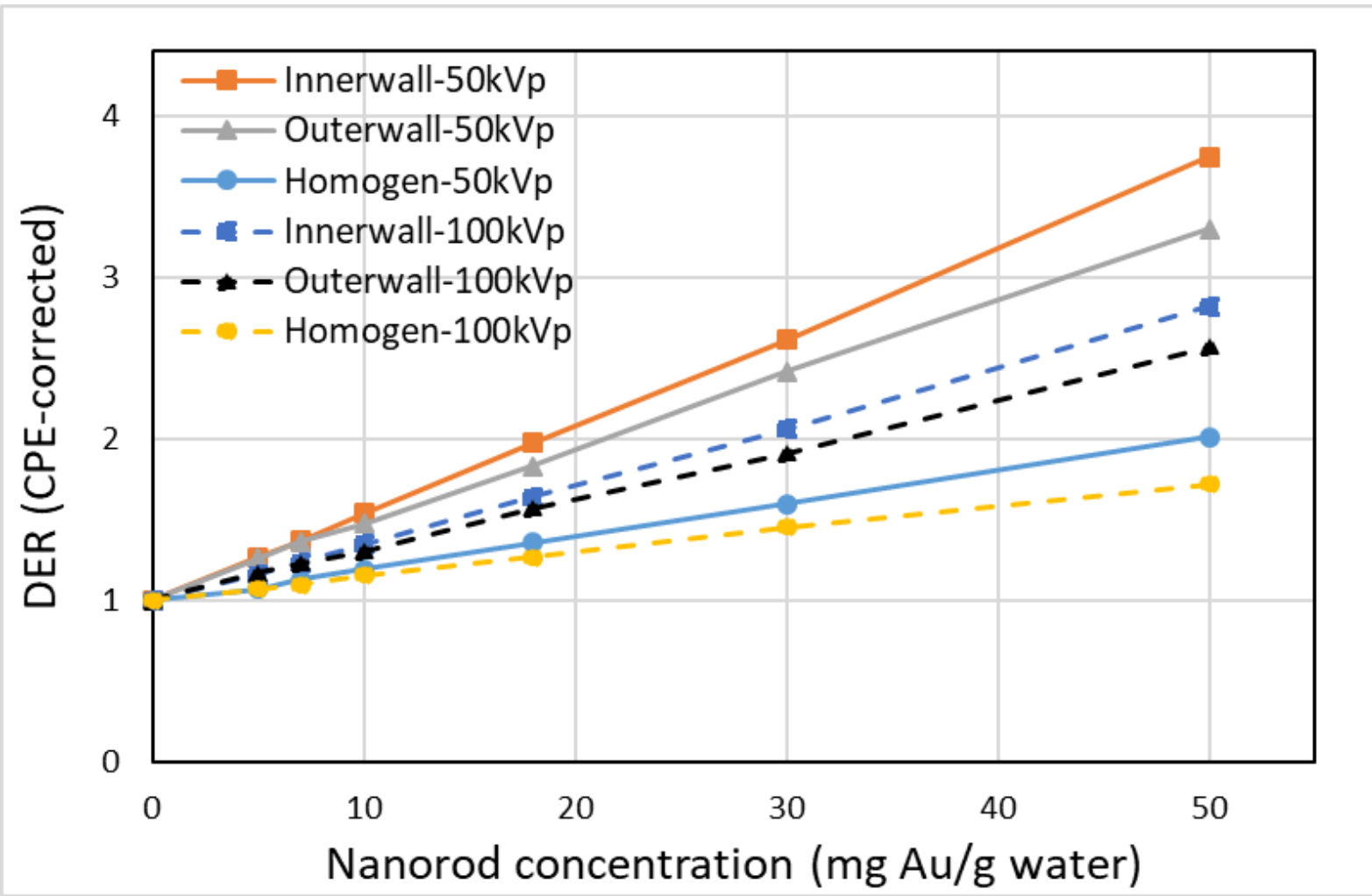

*Figure 5 Microvessel DER, averaged over the six endothelial cells, as a function of AuNR concentration for the three distributions at 50 and 100 kVp.*

The InnerWall+50 kVp combination consistently achieves the highest microvessel DER across all concentrations, reaching approximately 3.8 at 50 mg Au/g water, followed by OuterWall+50 kVp. The higher DER at 50 kVp compared to 100 kVp reflects the increased photoelectric absorption and production of low-energy secondary electrons. The InnerWall distribution's superior enhancement also aligns with the previous endothelial dose data.

Although our simulations explore AuNR concentrations up to 50 mg Au/g water, biological reality limits the deliverable nanoparticle doses due to cytotoxicity concerns. Preclinical mammalian studies report tumor-averaged gold concentrations up to 20-25 mg Au per gram of tumor, achieved safely after intravenous administration (López-Valverde et al., 2022; Xie et al., 2021). Multiple Monte Carlo simulation investigations have also used similar ranges, occasionally extending to 30 to 66 mg Au/g water (Koger and Kirkby, 2016; Martinov et al., 2023). These values appear to represent a biological threshold, beyond which the systemic clearance, organ accumulation, and toxicity risks increase. Thus, while our study simulates up to 50 mg Au/g water for physical dose-enhancement analysis, clinically relevant concentrations are likely to be between ~5 and 25 mg Au/g water. This again highlights the importance of combining physical modeling with targeted delivery strategies (e.g., vascular localization, surface functionalization, or intratumoral injection) to maximize radioenhancement in the tumor microenvironment within biologically feasible thresholds. Furthermore, while the 50 kVp spectrum produced the highest DERs, it is important to clarify that the clinical scope of these specific enhancement values is restricted to superficial and orthovoltage therapies, as the poor tissue penetration of kilovoltage X-rays precludes their use for deep-seated tumors.

It should also be noted that the uniform nanorod distributions used in this study within each scenario may not fully capture the physiological heterogeneity of tumor vasculature targeting. In reality, nanoparticles can aggregate in vivo, and aggregation status alters their tissue and cellular distribution (Keene et al., 2012). Irregular tumor vasculature also produces heterogeneous nanoparticle uptake and intratumoral distribution (England et al., 2015), and extravasation depends on nanoparticle shape (Jiang et al., 2024). These effects can lead to localized dose enhancement variations. Such clustering would be expected to increase microscale dose heterogeneity by producing higher peak at endothelial or perivascular dose near aggregates, and potentially reducing spatial uniformity of enhancement across the vessel segment. Nevertheless, assuming a uniform distribution in this model was a necessary first-order approximation to establish the fundamental physical relationship between gold concentration and dose enhancement.

### 3.2 Radial dose enhancement outside the vessel

Analyzing the radial distribution of radiation around tumor vasculature is essential for evaluating the therapeutic potential of nanoparticle-enhanced radiation therapy, as it outlines the spatial extent of enhanced radiation effects reaching the perivascular region, a critical tumor microenvironment containing 2-3 layers of tumor cells located within the first 15-20 µm of vessel walls (Nowosad et al., 2023). This region typically contains cancer stem cells responsible for tumor progression, therapeutic resistance, and metastatic spread. Radial DER directly influences the entire perivascular region (including the perivascular stem cell niche), as the extended dose enhancement provides a physical foundation for depositing localized energy into this critical region, which is known to host therapy-resistant cancer stem cells.

The CPE-corrected radial DER profiles, calculated within concentric cylindrical volumes extending up to 40 µm from the tumor microvessel surface under 50 and 100 kVp X-ray irradiation at a nanorod concentration of 18 mg Au/g water, are presented in Figure 6. The graphs reveal distinct trends for different distributions near the microvessel. Expectedly, the OuterWall distribution shows the highest DER near the microvessel surface for both X-ray spectra, followed by the InnerWall and Homogen distributions, with all exhibiting a sharp decline within the first ~10 µm. Beyond 10 µm, the DER for all distributions fall well below 1.05, indicating less than 5% physical dose enhancement. These findings are consistent with the limited range reported by Lin et al. (2015) for single-nanoparticle contributions, where the radial dose at 10 µm is at least $10^6$ times lower than at 10 nm. The present profiles give the DER of perivascular shells for three nanorod locations. Beyond ~10 µm, the DER of all distributions falls below 1.05. This threshold is important, as it defines the spatial boundary where localized physical dose enhancement from the nanoparticles becomes negligible. The pronounced dose enhancement observed near the microvessel surface, particularly for the InnerWall and OuterWall distributions, highlights the potential of vascular targeting to influence the proximal perivascular region, where cancer stem cells and metastatic cells commonly reside.

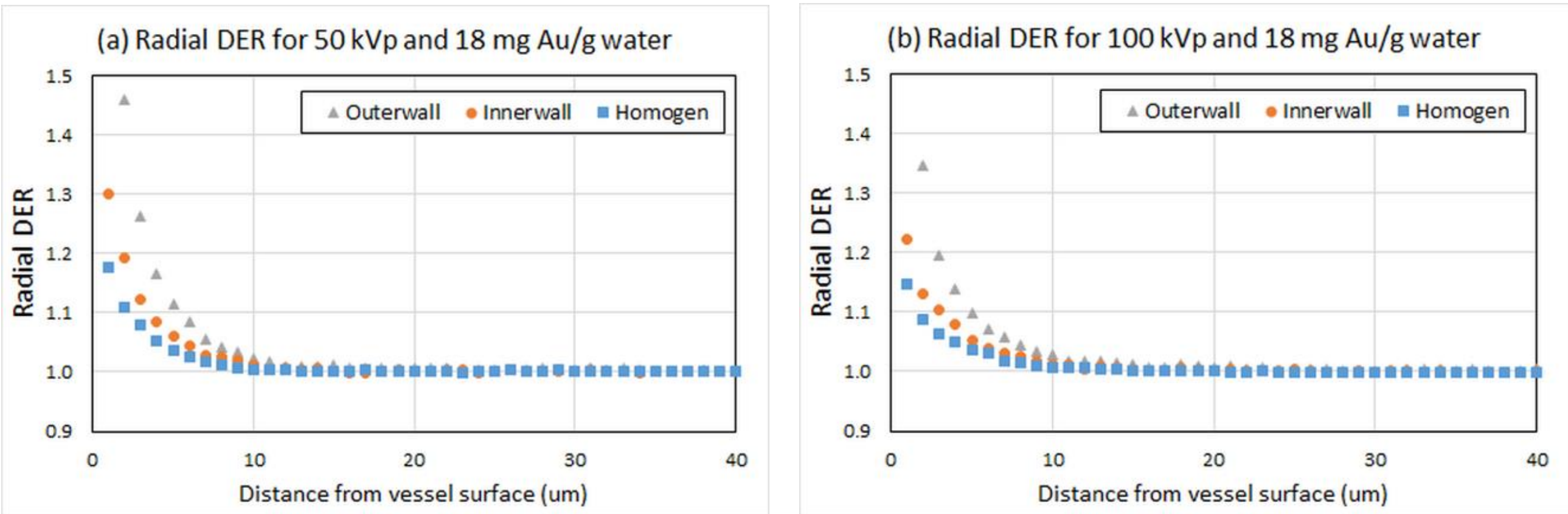

*Figure 6 Radial DER profiles for three distributions of 18 mg Au/g water gold nanorods under (a) 50 kVp, and (b) 100 kVp X-ray irradiation*

To further assess whether the finite vessel length considered in the simulations could affect the spatial extent of the reported dose enhancement, an independent analytical estimate based on the framework of Rabus and Thomas (2025) was performed and is presented in Supplementary Material 2. This approach evaluates dose contributions from AuNPs located upstream and downstream of the simulated vessel segment under charged-particle equilibrium conditions. Although these results are not directly comparable to the Monte Carlo data, since in simulations dose was scored in finite cylindrical shells and AuNRs were only present within the simulation geometry, the analysis consistently indicates an increase of the local dose enhancement (DER - 1) on the order of 10-20% for the wall-attached distributions when an effectively infinite vessel is assumed. Crucially, this additional contribution increases the magnitude of the proximal endothelial dose without extending the spatial range of enhancement beyond the immediate perivascular region, confirming that the reported DER profiles represent conservative estimates of localized vascular dose amplification.

However, although the DER becomes negligible beyond approximately 10 μm, the dose enhancement within the immediate perivascular region supports the concept of vascular targeting as a mechanism for amplifying the localized physical dose to endothelial cells, providing a dosimetric rationale for experimental observations of enhanced vascular damage. This observation highlights the potential of vascular-targeted AuNP radioenhancement as an effective strategy for tumors in which direct radiosensitization of tumor cells is limited by low nanoparticle uptake, hypoxia, or other biological barriers (Wu et al., 2022). Moreover, the dominance of short-range vascular effects may help explain discrepancies between experimental and theoretical radioenhancement studies, as *in vivo* experiments often capture indirect biological amplification mechanisms that are not fully represented in dose-based simulations (Dheyab et al., 2023; Yan et al., 2021).

While this study establishes a robust physical framework for vascular radioenhancement, specific modeling limitations must be acknowledged. First, the simulations assume a uniform distribution of AuNRs, whereas physiological delivery often results in nanoparticle aggregation and heterogeneous perivascular extravasation, which would likely create highly localized dosimetric profiles. Second, as noted in the analytical estimates, our framework does not fully account for the anisotropy of secondary electron emission. Finally, the reported DERs represent purely physical dose amplification. While track-structure simulation codes are highly capable of modeling downstream biological endpoints such as DNA damage and repair kinetics, the scope of the present study was intentionally restricted to physical dosimetry. Consequently, the ultimate radiosensitization efficacy *in vivo* will also depend on chemical, biological, and tumor microenvironment mechanisms (Liu et al., 2026) that extend beyond the physical dose parameters evaluated here.

## 4 Conclusion

This study quantified the impact of gold nanorod (AuNR) spatial localization within tumor vasculature on physical dose enhancement under kilovoltage X-ray irradiation using track-structure Monte Carlo simulations, establishing findings relevant to the clinical scope of superficial and orthovoltage radiotherapy. Three nanorod distributions, homogeneous lumenal, inner wall-attached, and outer wall-attached, were investigated in a segmented microvessel model with cell-resolved scoring of endothelial dose. To ensure physically meaningful dose enhancement ratios (DERs), all results were corrected for the lack of charged-particle equilibrium (CPE) inherent to confined beam microscale simulations.

Following CPE correction, AuNRs attached to the inner vessel wall consistently produced the highest microvessel-averaged DERs, reaching ~3.8 at 50 mg Au/g water and ~1.96 at the reference concentration of 18 mg Au/g water under 50 kVp irradiation. Dose enhancement increased linearly with AuNR concentration across the examined range, and 50 kVp X-rays yielded steadily higher enhancement than 100 kVp due to increased production of low-energy secondary electrons. Additional correction accounting for dose contributions from AuNRs located in adjacent vessel segments, representing an effectively infinite vessel, resulted in an increase of endothelial cell DERs by approximately 5-10%, without altering the relative trends between irradiation energies or nanoparticle distributions.

The simulations further show that physical dose enhancement extends several micrometers beyond the vascular wall into the proximal perivascular region, with the strongest enhancement confined to the first ~10 μm from the vessel surface. This spatially resolved characterization defines the physical dose conditions within endothelial cells and the immediately surrounding tissue that could drive vascular damage and amplified radiotherapy responses observed in experimental biological models. Overall, this work provides cell-resolved, CPE-corrected physical radioenhancement values for vascular-targeted AuNPs, and determines the influence of nanoparticle localization and irradiation energy on endothelial and perivascular dose. The results can serve as a physical basis for future radiobiological models of vascular radiosensitization.

**Acknowledgment:** Sunway University Studentship support received by Ali Taheri is acknowledged.

**Competing interests:** The authors declare no known competing interests.

**Declaration of generative AI technologies in the manuscript preparation process:** During the preparation of this work, the authors used Google Notebook LM in order to analyze the literature and Claude AI to improve language and readability. After using this tool, the authors reviewed and edited the content as needed and took full responsibility for the content of the publication

## Supplementary Material 1: Dose normalization and correction for lack of charged-particle equilibrium

This supplement describes the procedure used to correct dose enhancement ratios (DERs) obtained from confined-beam microscopic Monte Carlo simulations for the lack of charged-particle equilibrium (CPE). The macroscopic phase space generation simulation results were used to determine the photon fluence per primary particle incident on the microscopic geometry. Photons were scored in a planar scoring field (PSF1), and the photon fluence per primary fluence was calculated as:

$$\varphi_1 = \frac{N_1}{A_1} \times \frac{A_0}{N_0}$$

where $N_0$ is the number of primary particles simulated, $A_0$ is the area of the photon source, $N_1$ is the number of photon events scored in PSF1, and $A_1$ is the cross-sectional area of the PSF1 scoring volume.

The macroscopic dose per primary photon fluence under CPE conditions, $(D/\Phi)_1$, was calculated using the Excel templates from (Rabus *et al* 2019) for water at a depth of 2 mm. The values used were $6.31\times10^{-13}$ Gy cm$^2$ for 50 kVp irradiation and $4.17\times10^{-13}$ Gy cm$^2$ for 100 kVp irradiation.

The second set of microscopic simulations (tumor vessel irradiation) was then performed using a confined photon beam incident on the segmented microvessel geometry, both with and without gold nanorods (AuNRs). Due to the limited beam size, these simulations do not satisfy CPE. The dose contribution from the confined beam to endothelial cells and cylindrical scoring shells, per primary fluence, was calculated as:

$$(d/\Phi)_2 = d_2/N_2 \times A_2 \times \varphi_1$$

where $d_2$ is the dose scored in the microscopic simulation, $N_2$ is the number of primary particles, $A_2$ is the area of the photon source in the microscopic simulation, and $\varphi_1$ is the photon fluence ratio obtained above.

The additional dose contribution arising from AuNRs was obtained as:

$$(\Delta d/\Phi)_2 = (d_g/\Phi)_2 - (d_w/\Phi)_2$$

where $(d_g/\Phi)_2$ and $(d_w/\Phi)_2$ are the confined-beam dose contributions calculated with and without AuNRs, respectively. The dose enhancement ratio was then calculated as:

$$DER = 1 + (\Delta d/\Phi)_2/(D/\Phi)_1$$

This formulation yields CPE-corrected DER values that isolate the physical dose enhancement due to AuNRs while accounting for the confined-beam nature of the microscopic simulations. Contributions from AuNRs located outside the simulated vessel segment (i.e., upstream or downstream) are not included here and are addressed separately in Supplementary Material 2.

## Supplementary Material 2: Analytical estimation of dose contributions from gold nanorods outside the simulated vessel segment

To estimate the effects of AuNRs in the blood vessel or near the endothelium outside the geometry considered in the simulations, the approach proposed recently in (Rabus and Thomas 2025) was adapted and implemented in a GDL script for numerical integration of the weighting factors for the dose distribution around a single gold nanoparticle. The corresponding data for a 100 nm spherical AuNP irradiated by 50 kVp and 100 kVp X-rays were taken from (Li *et al* 2024), using the difference of the simulation results with and without AuNP as the contribution to the dose from an AuNP undergoing an ionizing interaction. The expected number of interactions per dose were calculated following a development of the approach presented in (Rabus *et al* 2019) in which only ionizing interactions are considered.

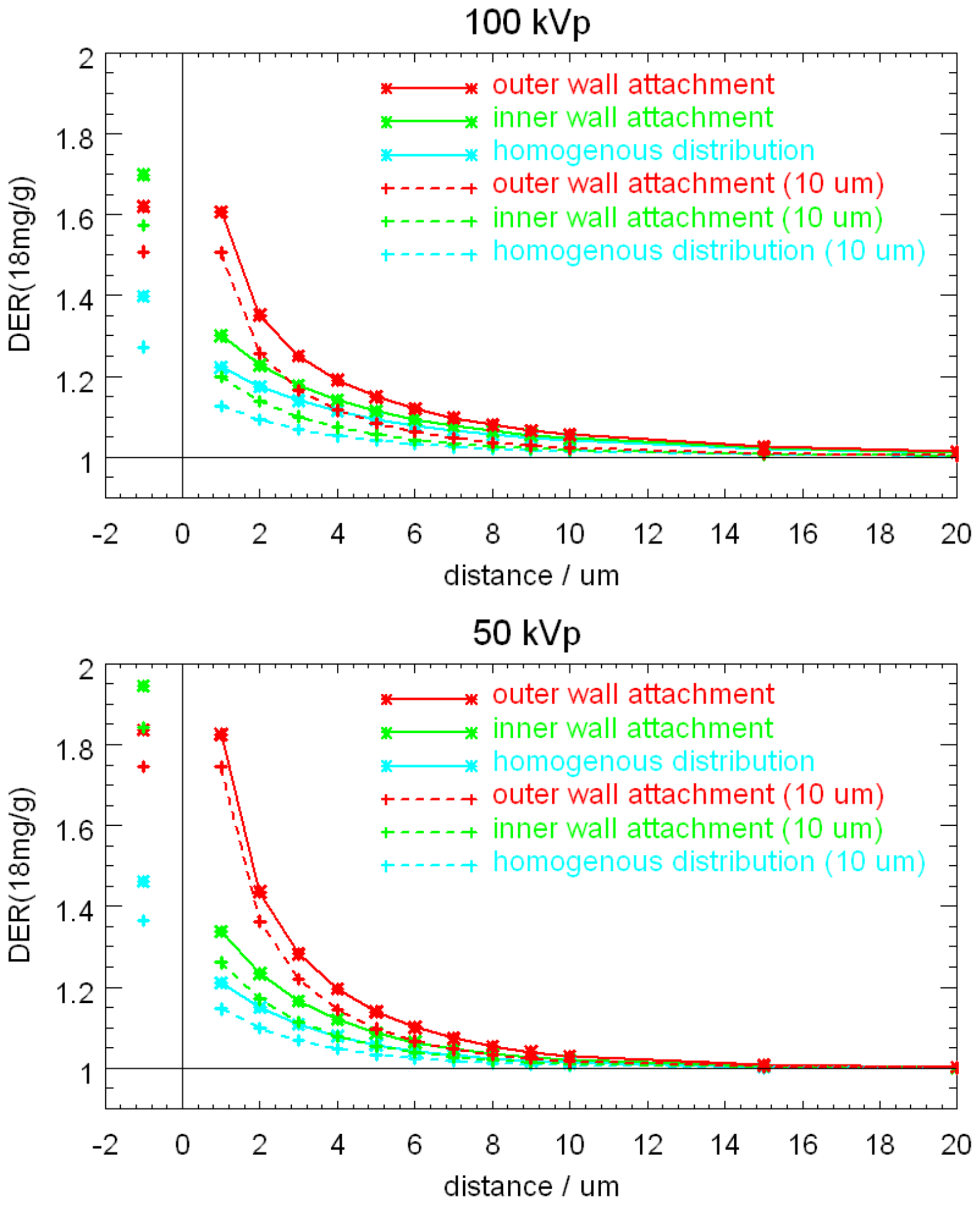

*Supplementary Fig. S1: Dose enhancement in and around the vessel walls under CPE conditions when only the AuNPs inside the geometry of Fig. 2b are considered (plus signs) or an infinite distribution is assumed (asterisks). The x-axis shows the distance from the outer wall of the vessel in µm.*

The results shown in Supplementary Fig. S1 correspond to a AuNR distributions with a gold mass fraction of 18 mg Au/g water within the vessel. The data represent the dose enhancement at the center of endothelial cells and at the centers of radial scoring shells surrounding the vessel segment. The plus symbols indicate results obtained when only AuNRs within the simulation geometry shown in Fig. 2b are considered, whereas asterisks indicate results obtained when AuNRs are assumed to be present at the same concentration in the upstream and downstream portions of the vessel.

The inclusion of AuNRs outside the simulated vessel segment results in an increase in dose enhancement (DER − 1) of approximately 10-20% for the wall-attached distributions and a larger relative increase for the homogeneous distribution. These values are not fully comparable to the Monte Carlo simulation results, as the simulations scored dose in cylindrical shells of 10 µm length whereas here the dose was determined in the center of these shells. However, it is expected that the relative difference between AuNRs only inside the vessel segment and also outside it, are comparable.